# Feature learning of virus genome evolution with the nucleotide skip-gram neural network

Hyunjin Shim[1,2,3]

[1]*Stanford University, Stanford, California, USA*
[2]*Ecole Polytechnique Fédérale de Lausanne (EPFL), Lausanne, Switzerland*
[3]*Swiss Institute of Bioinformatics (SIB), Lausanne, Switzerland*

EPFL SV-DO
SV 3807 (Bâtiment SV)
Station 19
CH-1015 Lausanne
Switzerland
hyunjin.shim@epfl.ch

## Abstract

Recent studies reveal even the smallest genomes such as viruses evolve through complex and stochastic processes, and the assumption of independent alleles is not valid in most applications. Advances in sequencing technologies produce multiple time-point whole-genome data, which enable potential interactions between these alleles to be investigated empirically. To investigate these interactions, we represent alleles as distributed vectors that encode for relationships with other alleles in the course of evolution, and apply artificial neural networks to time-sampled whole-genome datasets for feature learning. We build this platform using methods and algorithms derived from Natural Language Processing (NLP), and we denote it as the nucleotide skip-gram neural network. We learn distributed vectors of alleles using the changes in allele frequency of echovirus 11 in the presence or absence of the disinfectant ($ClO_2$) from the experimental evolution data. Results from the training using a new open-source software TensorFlow show that the learned distributed vectors can be clustered using Principal Component Analysis and Hierarchical Clustering to reveal a list of non-synonymous mutations that arise on the structural protein VP1 in connection to the candidate mutation for $ClO_2$ adaptation. Furthermore, this method can account for recombination rates by setting the extent of interactions as a biological hyper-parameter, and the results show the most realistic scenario of mid-range interactions across the genome is most consistent with the previous studies.

## 1 Introduction

In evolutionary biology, statistical models are used to investigate the evolutionary processes that generated the vast diversity of life. Our current understanding of evolution stems from the work of many biologists that derived theoretical models from the observations in nature, such as selection, genetic drift, migration, speciation (Haldane 1927, Fisher 1930, Wright 1931, Kimura 1968, Charlesworth 1993). However, these models are established under some assumptions that simplify the underlying principles, as most biological processes are dynamic and complex. For example, the Wright-Fisher model is one of the fundamental population genetic models, which represents the process of genetic drift as a binomial sampling of 2N gene copies between generations in an idealized population of N (Fisher 1930, Wright 1931). This idealized population has no overlapping generations, and each gene copy is independently drawn to the next generation at random in a fixed population. Despite these simplifying assumptions, the Wright-Fisher model is still widely used in population genetic methods, for instance in a recent analysis of forward processes in time-sampled datasets (Foll et al. 2014a,b, Shim et al. 2016). Some assumptions of the model are valid in most applications, as the model has been proved to give intuitive approximations of more complex real cases (Ewens 2004). However, the assumption of independent alleles is problematic in many cases, particularly as advances in molecular genetics unveil how alleles interact in intricate networks to produce unexpected outcomes, even for viruses with comparatively small and simple genomes (Burch et al. 2003, Michalakis et al. 2004, Kryazhimskiy et al. 2011, Ibeh et al. 2016). Furthermore, advances in sequencing technologies generate datasets such as whole-genome sequences where the relations between alleles can potentially be investigated, particularly in time-serial cases (Malaspinas et al. 2012, Foll et al. 2014a,b, Shim et al. 2016). In order to investigate these complex interactions, we propose to represent alleles as distributed vectors that encode for relationships with other alleles in the course of evolution, rather than as discrete entities as in the case of conventional models like the Wright-Fisher model. This novel concept of representing alleles as distributed vectors is inspired from Artificial Neural Networks and Natural Language Processing, and we use a model-free approach to learn these distributed vectors of alleles directly from genetic sequence data.

Artificial Neural Networks (ANNs) are biologically-inspired computing elements that can be interconnected to process and learn multiple levels of representation from external input information such as sound, image or characters. ANNs are becoming increasingly popular in many fields, as they provide a flexible framework where learned features are easy to adapt and learn, without manual over-specification of features like other machine learning techniques. Since 2010, multilayer neural networks started outperforming other machine learning techniques in speech and vision with the availability of big data and faster machines (Schmidhuber, 2015). There has been a recent interest to apply ANNs in medicine and biology due to the exponential growth in data production with technological advances, such as in structural biology, regulatory genomics, drug discovery, and cell imaging (Angermueller et al. 2016; Mamoshina et al. 2016). However, in the evolutionary context, only a few attempts have been made to apply ANNs in classic population genetic problems such as inference of selection and demography from natural populations (Sheehan and Song, 2013). Here, we use Artificial Neural Networks in the evolutionary framework to exploit their ability to learn from high-dimensional biological data without explicit programming or modeling or prior knowledge through feature learning. The algorithms and training methods are derived from the Skip-gram Neural Network model in Natural Language Processing (NLP) that aims to represent the meaning of a word in distributed rather than discrete representations (Mikolov et al. 2013a,b). As opposed to discrete representations where there is no natural notion of similarity, distributed representations capture the context of a word over a large vector space. Distributional similarity-based representations define a word by means of its neighbors, where a word vector is trained using ANNs to predict neighboring words given a center word in the skip-gram neural network model. In a similar approach, we attempt to represent alleles by distributional similarity-based representations by predicting between a center allele and its neighboring alleles, with the aim of deciphering genetic interactions between these alleles along the course of a time-serial experimental evolution.

Viruses are the most straightforward experimental models to study the interactions between alleles, and they have been investigated through theoretical modeling and empirical analysis using site-directed mutagenesis and mutation-accumulation experiments (Sanjuan et al. 2004, 2005, Elena et al. 2010, Rokyta et al. 2011, Lalic et al. 2012). These studies reveal that virus genomes evolve under complex patterns of genetic interactions such as epistasis and clonal interference to shape the architecture of fitness. Here, we learn these interactions between alleles as biological features through neural networks from mutation-accumulation experiments, where no model assumptions are needed as in the previous investigations. We use population-level whole-genome sequences from the time-sampled experimental evolution of echovirus 11 in the presence or absence of the disinfectant treatment to train this ANN platform, where features to learn are distributed vector representations of the alleles. After training, we cluster these alleles by similarity in the evolutionary trajectory using Principal Component Analysis and Hierarchical Clustering, and we compare these clusters of potential genetic interactions with the previous investigation and further discuss the advantages of representing alleles as distributed vectors rather than discrete entities.

# 2 Methods

## 2.1 The Nucleotide Skip-gram Neural Network

The skip-gram model uses distributed vector representations of words to predict for every center word its context words (Mikolov et al. 2013a,b). The model uses neural networks for learning these word vectors from large datasets, and we apply the analogous neural network architecture as shown in Figure 1 to find distributed vector representations of alleles to predict between every center allele and its nearby alleles from time-sampled allele frequency datasets. Here, an allele is defined as a nucleotide that increases in frequency more than the sequencing error between two sampling time points. Given a nucleotide sequence of training data $a_1$, $a_2$,…, $a_T$, the objective function is to maximize the average log probability:

$$\frac{1}{T}\sum_{t=1}^{T}\sum_{-w\leq j\leq w, j\neq 0}\log p(a_{t+j}|a_t)$$

where $w$ is the window size of the nearby alleles that defines the extent of interaction. The skip-gram model defines $p(a_{t+j}|a_t)$ using the softmax function as following:

$$p(a_O|a_t) = \frac{\exp\,(u_{a_O}^T v_{a_t})}{\sum_{m=1}^{M}\exp\,(u_m^T v_{a_t})}$$

where $a_o$ are the nearby alleles, $M$ is the size of the set of alleles in the dataset, and $v_a$ and $u_a$ are the "input" and "output" vectors that represent the center and nearby distributions for each allele, respectively. Here, the set of alleles comprises of all the alleles in the genome that increase in frequency more than the sequencing error between two sampling time points, which can be newly arising mutations (Single Nucleotide Polymorphism: SNP) or standing variation. As shown in Figure 1, the nucleotide skip-gram neural network is composed of one hidden layer computing the projection and one output layer computing the softmax function. We hereby denote distributed vector representations of alleles as 'allele embeddings', as analogous to 'word embeddings' in Natural Language Processing.

## 2.2 Noise-Contrastive Estimation

In practice, an approximate of the full softmax is computed by Noise-Contrastive Estimation (NCE), as normalizing each probability at every training step is computationally expensive (Mnih, Kavukcuoglu 2013). The score of the similarity measure between the center allele $a_t$ and the nearby alleles $a_o$ is given as a dot product. The dot product, $\text{score}(a_o, a_t)$, is then converted to a probability using a softmax function, and the maximum likelihood (ML) is used to maximize the probability of the nearby alleles ($a_o$) given the center allele ($a_t$) for the nucleotide skip-gram model:

$$\mathrm{P}(a_o|a_t) = \mathrm{softmax}(\mathrm{score}(a_o, a_t))$$

$$\mathrm{J_{ML}} = \log \mathrm{P}(a_o|a_t) = \mathrm{score}(a_o, a_t) - \log(\sum_{a'} \exp\{\mathrm{score}(a', a_t)\})$$

where $a'$ is the set of alleles increasing above the sequencing error. However, computing the score for all other $a'$s in the current center allele $a_t$ at every training step is expensive, and a full probabilistic model is not necessary for feature learning. The nucleotide skip-gram model uses a binary classification object to discriminate the nearby alleles $a_o$ from $k$ noise alleles $\widetilde{a_o}$ , in the same center allele:

$$\mathrm{J_{NEG}} = \log \mathrm{Q_\theta}(\mathrm{D} = 1|a_o, a_t) + \mathrm{k}\mathbb{E}_{\widetilde{a_o}\sim \mathrm{P_{noise}}}[\log \mathrm{Q_\theta}(\mathrm{D} = 0|\widetilde{a_o}, a_t)]$$

where $\mathrm{Q_\theta}(\mathrm{D} = 1|a_o, a_t)$ is a binary logistic regression of having the nearby alleles $a_o$ in the center allele $a_t$ in the dataset $D$, calculated in terms of the learned allele embeddings $\theta$. The objective is maximized when high probabilities are assigned to the correct alleles over $k$ noise (contrastive) alleles whose expectation is approximated by Monte Carlo sampling from the noise distribution $P_{noise}$ (typically the uni-gram distribution). With the NCE, the objective function computation now scales with the number of noise alleles $k$ instead of all alleles $a'$ in the set. For our datasets, a simplified variant of Noise-Contrastive Estimation called Negative Sampling is used, in which only samples are used instead of the numerical probabilities of the noise distribution to approximate the full softmax.

# 3 Implementation

## 3.1 Biological model and training data

We used the nucleotide skip-gram neural network in unsupervised setting to train the distributed vector representations of the alleles using the data from the experimental evolution of echovirus 11 under the presence or absence of the disinfectant, $ClO_2$ (Zhong et al. 2017). In this experiment, a wild-type (WT) population of echovirus 11 was repeatedly exposed to chlorine dioxide ($ClO_2$), which is a highly effective disinfectant that inactivates a broad range of waterborne viruses. Following inactivation, the surviving viruses were passaged onto a monolayer of BGMK cells for regrowth. For each disinfection stage, the virus population was subjected under $ClO_2$ concentration so as to reduce the inactivation rate constant by approximately 50%. After 20 cycles of disinfection-regrowth in 2 replicates (EA and EB), the whole genomes of the WT and evolved virus populations were sequenced with Next Generation Sequencing (Illumina HiSeq 2500). As a control experiment, the same WT viruses were subjected to 10 cycles of bottleneck events by dilution without being exposed to $ClO_2$ (NEA and NEB) and regrowth in cell culture, followed by the whole-genome sequencing of the evolved virus population.

The training datasets were generated from the whole-genome datasets using the increase in the minor allele frequencies between the two sampling time points (WT and evolved states). Here, a nucleotide sequence of minor alleles $a_1$, $a_2$,..., $a_T$, is defined as the Single Nucleotide Polymorphism (SNP) or standing variation with the highest frequency in the evolved populations. As the datasets were pool-sequenced, we reconstructed the difference in the raw reads of each allele between two sampling time points by simulating each site in the virus genome as the binomial distribution with the probability of success as the increase in the allele frequency, $f$:

$$\Pr(X = k) = \binom{n}{k} f^k (1 - f)^{n-k}$$

where $n$ is the number of raw reads and $k$ is the number of minor alleles at each site. The genome-wide average of the coverage depth is approximated by taking the mode of the raw read depths across the virus genome at each sampling time point as shown in Table 1 and Figure S1. We only retained the minor alleles whose increase in frequency was above the sequencing error of the Illumina sequencer, which was previously validated to be 1% from the experimental study (Zhong et al. 2017). The combined number of the retained minor alleles in the exposure replicates (EA and EB) and control replicates (NEA and NEB) is 86 and 72, respectively.

### 3.2 Training neural networks with TensorFlow

Skip-gram models are a generalization of *n*-grams that model sequences, in which components may be skipped over rather than being in consecutive order. Here, we have a simple neural network with a single hidden layer to train, where the weights of the hidden layer are allele embeddings that are the features to be learned. These weights are initialized randomly from a uniform distribution of [-1,1]. For this feature learning, we trained the neural network in TensorFlow (Version 1.2.1) to optimize the probability for every allele in our datasets of being the nearby allele given the center allele, as illustrated with TensorBoard in Figures 2 and S2. From our datasets of the allele sequences, center and nearby allele pairs are randomly chosen as [input, output] within the given window size $w$ as described in the Methods. The NCE training objective defined above is optimized with Stochastic Gradient Descent (SGD) using mini batches, where the batch size of 128 was used in our optimization step. The output vectors are the probabilities of all alleles in our datasets being the nearby allele for the chosen center allele, optimized using the mini batches of the generated [input, output] pairs.

### 3.3 Hyper-parameters and Bio-parameter optimization

In our nucleotide skip-gram neural network, we have three hyper-parameters to consider: the dimension $N$ of the distributed vectors in the hidden layer, the number of negative samples $k$ per positive sample, and the learning rate $r$. After several optimization runs, we chose the simplest hyper-parameters among the test values as 128 neurons in the hidden layer, 16 negative samples, and the learning rate of $1 \times 10^{-2}$, respectively.

Here, we define the window size $w$ as a bio-parameter since its value depends on the architecture of a virus genome. The window size designates how far a center allele is assumed to interact with its nearby alleles in the genome, potentially representing the extent of biological factors such as epistatic interaction or clonal interference. Thus, the window size must depend on the recombination rate of the virus model under investigation. The recombination of echovirus 11 has not yet been measured directly, but its closely related species, Poliovirus, is known to have a recombination rate of $3\text{-}7 \times 10^{-6}$ per nucleotide per generation (Reiter et al. 2011). For the genome size of $7 \times 10^{3}$ in echovirus 11, a similar recombination rate amounts approximately to $2\text{-}5 \times 10^{-2}$ recombination events per generation. Thus, we considered following three cases where the center allele amid the set of $M$ alleles is linked to:

1) only the immediate neighbors ($w$=1),
2) quarter of the neighbors on either side ($w = M/4$ if $M$ is even, or $w=(M-1)/4$ if $M$ is odd),
3) half of the neighbors on either side ($w = M/2$ if $M$ is even, or $w=(M-1)/2$ if $M$ is odd).

These three cases represent respectively, 1) a high recombination rate that the center allele is linked to only neighbors located nearby, 2) a moderate recombination rate that the extent of interaction reaches approximately half of the genome, and 3) a low recombination rate that the extent of interaction reaches almost the entire genome.

# 4 Results

## 4.1 Visualization with TensorBoard

After $10^5$ training steps as shown in Figure S3, the distributed vectors of the alleles were visualized using TensorBoard for the exposed and non-exposed experiments (Figures S4-S6). The graphs from TensorBoard show the cosine distances between the allele embeddings in the original space learned from the nucleotide skip-gram neural network. Each point is indexed to the nucleotide position in the genome, and the allele of interest in the exposed population (P129Q denoted as Position 2844) and the non-exposed population (H215N denoted as Position 3101) is highlighted. According to the previous literature, these two non-synonymous mutations are potential drivers that give rise to the new functional adaptation to the echovirus under the given challenging conditions (Stuart et al. 2002, Rezaikin et al. 2009, Zhong et al. 2017). Zhong et al. (2016) previously demonstrated that $ClO_2$ impairs host binding, thus the evidence of resistance to $ClO_2$ in these experiments indicates the echovirus is able to evolve an enhanced binding mechanism that can counteract the disinfectant acting on the viral genome. Furthermore, Zhong et al. (2016) postulate that these adaptive mutations are likely to be located on the structural proteins of VP1 and VP2, as these specific mutations allow echovirus to use an alternative co-receptor that strengthens virus binding (Stuart et al. 2002). As the candidate allele on VP2 was already present as a major allele in the WT at position 139, the candidate mutation for enhanced host binding was deduced to have arisen on VP1 - as K259E in the exposed population and as H215N in the non-exposed population. These mutations may be one of the factors that render the echovirus with replicative advantages compared to the WT under the challenges of repeated bottlenecks. Furthermore, another mutation at P129Q is important under the presence of $ClO_2$ in the exposed population, as it causes the substitution of a $ClO_2$-reactive amino acid (proline) to a $ClO_2$-stable amino acid (glutamine), increasing the $ClO_2$ endurance of the protein capsid (Zhong et al. 2017).

## 4.2 Correlation analysis with Principal Component Analysis

Principal Component Analysis (PCA) is a statistical technique that converts data of correlated variables to linearly uncorrelated variables called principal components. PCA was carried out on the learned allele embeddings for the three window sizes ($w$=1, $w$=$M$/4, $w$=$M$/2) using TensorBoard (Figures S4-S6). The total variance described with the first three PCA components is summarized in Table S1. The pair-wise correlation matrices were generated by calculating the cosine distances between the first three PCA components of the allele embeddings, and they were visualized in the genomic position order as correlation maps in A, B of Figure S7-S9. In order to mine for hidden patterns, we applied hierarchical clustering to each matrix to obtain agglomerative correlation maps of the allele embeddings as shown in C, D of Figure S7-S9. Each correlation map contains one empty box that is designated for all zeros in the data. As shown in the figures of hierarchical clustering (C, D of Figure S7-S9), it is notable that the correlation matrices display clear clustering of the alleles by similarity (blue) as well as by dissimilarity (red). The patterns of similarity and dissimilarity are more distinct in the exposed experiments, indicating the evolution of echovirus 11 was more directional in the presence of the disinfectant $ClO_2$ as compared to that of the control experiments.

The hierarchical clusters of the allele embeddings were compared to the previous results (Zhong et al. 2017), where the mutation clusters were identified by calculating Pearson correlation coefficients between two mutations using their allele frequencies and setting a cluster threshold of 0.95. The Pearson correlation method in this study was limited to pair-wise correlations in the allele frequencies between two mutations from a manually specified set, and it was unable to take into account of the complete datasets (for example, differences in replicates). We chose the cluster size of seven as used in the previous analysis, and the results from the two approaches were compared as shown in Table 2 and Table S2. In Table 2, the clusters containing the alleles of interest (P129Q and H215N) in the exposed and non-exposed populations are shown in detail. The comparison reveals that the allele clusters from the nucleotide skip-gram neural network with the window size as *M*/4 are the most consistent with the Pearson correlation clusters by Zhong et al. (2017), for both the exposed and non-exposed populations. These clusters with the window size as *M*/4 have the highest number of the overlapping alleles to the previous list and they all lie within the protein-coding genome. It is intriguing that this window size of *M*/4 represents a moderate recombination rate, as the alleles are assumed to interact between the quarter of the neighboring alleles on either side that are evenly distributed along the genome (Figure S10). Given the recombination event of $2\text{-}5 \times 10^{-2}$ per generation (Reiter et al. 2011), the case with a moderate recombination rate where the extent of interaction reaches approximately half of the genome during the evolution is the closest representation of the echovirus biology.

## 4.3 Virus protein evolution

We further analyze the most realistic scenario of the allelic interactions within the window size of *M*/4. Figure 3 shows that the correlation map of the allele embeddings can be arranged in positively correlated clusters using hierarchical clustering, and these clusters also show strong patterns of negative correlation as well. This result is consistent with the fact that when alleles are similar by distributed vector representations, they should also display a similar pattern of dissimilarity. This indicates that a cluster of alleles that increases in frequency in the similar context may behave antagonistically in a collective manner to a cluster of alleles of another context – however, biological interpretations to this result should be investigated further with experimental validations.

For the exposed population, the cluster that contains the candidate mutation P129Q that enhances host binding in the presence of $ClO_2$ (Zhong et al. 2016, 2017) is highlighted on the structural protein of VP1 in Figure 4. Most of the mutations on the virus structural proteins in this cluster are overlapping with the list from Zhong et al. (2017), except for the two mutations (T2849A, C2850A) on the VP1 protein. Interestingly, all but one of the mutations on VP1 in this cluster are non-synonymous, supporting the hypothesis of the important role of VP1 in adapting to the disinfectant. Furthermore, this cluster contains another candidate mutation (K259E) that was previously identified as contributing to adaptation, and this finding suggests a potential epistatic interaction between these two candidate mutations, P129Q and K259E, during the course of experimental evolution. Contrarily to the list from Zhong et al. (2017), this cluster also contains additional mutations from the nonstructural proteins, such as the proteins related to the viral polymerase (3C and 3D) and NTPase (2C). However, all mutations but one are synonymous, which indicates that this interaction may be an artifact of genetic drift or linkage rather than these mutations on the nonstructural proteins playing an adaptive role in the challenging environments.

For the non-exposed population, the candidate mutation that gives a fitness advantage to the wild type is H215N on the VP1 protein. We also compare the cluster containing this mutation, which may help usage of an alternative receptor binding, to the previous result (Zhong et al. 2017). In contrary to the list of Zhong et al. (2017) with only two mutations on the VP1 protein, the cluster of the nucleotide skip-gram neural network contains several mutations from both the structural and nonstructural proteins. Besides the two mutations on VP1 previously identified, there is a few non-synonymous mutations from the viral polymerase proteins (3A and 3D) (Lin et al. 2009) which are to be investigated further empirically whether their clustering with the candidate mutation can potentially result from genetic interactions.

## 5 Discussion

We present an application of artificial neural networks that learns biological features from time-sampled datasets of virus genome evolution. This application uses methods and algorithms derived from Natural Language Processing, such as the skip-gram model and Noise-Contrastive Estimation, to learn distributed vector representations of the alleles that increase in frequency above the sequencing error during the time-serial experimental evolution. To the best of our knowledge, this is the first attempt to represent alleles as distributed vectors instead of discrete entities as in conventional evolutionary models, enabling the relationships between these alleles to be encoded in a continuous vector space of low dimension. We learn these features through the neural networks by predicting for every center allele its neighboring alleles from the changes in allele frequency, and the data application using the time-sampled whole-genome sequences of echovirus 11 was carried out. The results show that the alleles rising above the sequencing error can be represented effectively as distributed vectors from genetic sequence data, as compared to being represented as discrete and independent entities in most classic population genetic models. Distributed vector representations of alleles learned from the nucleotide skip-gram neural network have the advantage of incorporating the information from neighboring alleles entirely from input data without model assumptions. Using Principal Component Analysis, the pair-wise correlation map between these allele embeddings is generated and arranged by agglomerative hierarchical clustering, which unveils the similarity in the evolutionary trajectory between these alleles. In comparing with the previous result by Zhong et al. (2017), the clusters with the window size of $M/4$ are the most plausible, representing a moderate recombination rate which is also consistent with the known echovirus biology. Furthermore, a few non-synonymous mutations are identified to have had evolved similarly to the candidate mutations of adaptation, and the question of whether evolution processes (genetic drift, selection) or genetic interactions (epistasis, clonal interference) are responsible for this pattern is of interest to future investigations.

The advantages of the nucleotide skip-gram neural network model include the absence of manual over-specification and model assumptions as needed in the previous studies of genetic interactions. Furthermore, this neural network platform has the potential to be applied to larger and more complex datasets, such as for organisms with bigger genomes or for data from natural populations. The nucleotide skip-gram neural network has a simple architecture of one hidden layer that minimizes computational complexity, which makes the platform ideal for much bigger datasets such as human genomes with approximately 3 billion base pairs. The caveat of this current data application is that the time-sampled whole-genome virus datasets from this experimental evolution are actually 'too small' for the capacity of neural networks, potentially leaving the results under-trained and sub-optimal. For future investigations, this method can be applied to time-sampled datasets from human cancer cells in order to decipher the interactions between the mutations that arise during the course of cancer evolution, or to spatial datasets from natural populations of humans or drosophila to decipher spatial rather than temporal interactions between the alleles of interest. Furthermore, biological factors such as recombination rate can be represented in more accurate ways to investigate whether the patterns of interaction are produced by deterministic or stochastic evolutionary forces. Lastly, technological factors such as a larger number of sampling time points and raw sequences can also be incorporated for future studies.

## Acknowledgements

We thank Serafim Batzoglou, Chip Huyen, Minwoo Sun, and Diego Marcos Gonzalez for helpful discussions. This work was supported by the Swiss National Science Foundation [grant number P1ELP3_168490 to H.S.]; and Firmenich EPFL-Stanford research exchange program.

## Tables

**Table 1.** Mode of coverage depth in the whole-genome sequencing of echovirus 11 with Illumina HiSeq 2500

| | **Exposed (E)** | **Non-Exposed (NE)** |
|---|---|---|
| **Replicate A** | 28994 | 25937 |
| **Replicate B** | 21373 | 21797 |

**Table 2.** Allele clusters identified by two approaches: (A) Pearson correlation coefficient (Zhong et al. 2017), and (B) Skip-gram neural network. For the nucleotide skip-gram neural network, the order of the alleles represents the order of similarity (see Table S2 for a complete list of clusters).

| **1) Cluster<br>2) Biological Function** | **(A)** | **(B)** | | |
|---|---|---|---|---|
| | **Exposed & Non-exposed** | **Exposed** | | |
| | | ***w*=1** | ***w*=*M*/4** | ***w*=*M*/2** |
| **1) II<br>2) Enhances host binding in the presence of $ClO_2$** | VP1:A2835G:K126R<br>VP1:C2844A:P129Q<br>VP1:T2849A:S131T<br>VP1:C2850A:S131Y<br>VP1:C3162T:T235I<br>VP1:A3170G:M238V<br>VP1:A3233G:K259E | VP1:A2854C:R132<br>3D:T6190G: H80Q<br>VP1:C3103A:H215Q<br>3D:T7240G:Y430*<br>3C:A5634C:N78T<br>3D:A7250C:I434L<br>VP1:C2844A:P129Q<br>VP1:T2849A:S131T<br>3A:T5203C:V45<br>:G7383A: | 3D:T5964C:F5S<br>3D:A7250C:I434L<br>VP2:T1660C:D234<br>VP3:A1761G:N6S<br>3C:A5893T:G164<br>3D:C6745T:Y265<br>VP1:C2632T:S58<br>2C:A4552G:L157<br>VP1:C2844A:P129Q<br>3D:G6409A:P153<br>VP1:A3233G:K259E<br>VP1:A3170G:M238V<br>VP1:C3103A:H215Q<br>VP1:A2835G:K126R<br>VP1:C3162T:T235I | VP1:A3170G:M238V<br>3D:G7246A:E432<br>3C:C5818G:N139K<br>VP1:A2835G:K126R<br>3D:T6006C:M19T<br>:G7383A:<br>VP2:C1666T:S236<br>3C:A5893T:G164<br>3A:T5323C:F85<br>VP1:C2844A:P129Q<br>VP1:A3233G:K259E<br>VP1:T2849A:S131T<br>3C:T5788A:G129<br>VP1:C3103A:H215Q<br>2C:G4384A:V101 |
| | **Exposed & Non-exposed** | **Non-exposed** | | |
| | | ***w*=1** | ***w*=*M*/4** | ***w*=*M*/2** |
| **1) V<br>2) Helps usage of an alternative receptor binding** | VP1:A2937T:Y160F<br>VP1:C3101A:H215N | 3C:G5710C:A103<br>3D:C6991T:T347<br>VP1:C3101A:H215N<br>2C:C4519T:L146<br>2C:T4666C:S195<br>VP2:C1008T:T17I<br>2A:C3367T:Y11<br>2C:C4454T:L125<br>VP1:C3285T:T276I<br>VP1:G2521A:G21<br>:G7346A:<br>3D:G7246A:E432<br>3D:C7249T:F433 | VP2:C1210T:D84<br>VP3:T1831C:D29<br>VP1:C3101A:H215N<br>2B:C3841T:N19<br>2C:C4546T:Y155<br>3D:C6991T:T347<br>VP1:A2937T:Y160F<br>3D:C6085T:N45<br>3A:A5306G:I80V<br>3D:A6989G:T346A<br>VP2:C1243T:N95<br>VP3:A1761G:N6S | VP2:C1008T:T17I<br>VP1:C3101A:H215N<br>VP1:A2937T:Y160F<br>3D:A6989G:T346A<br>3D:C6991T:T347<br>:G659A:<br>2C:C4546T:Y155<br>3A:A5306G:I80V<br>VP3:A1761G:N6S<br>3A:T5323C:F85 |

**Protein : Nucleotide position with the major and minor alleles : Nucleotide position within the protein with the original amino acid (and the substituted amino acid for non-synonymous mutations)**
*** indicates STOP codon**

**Figure 1.** Architecture of the nucleotide skip-gram neural network used to train distributed vectors of alleles, with one linear hidden layer (Projection) of N neurons and one output layer (Softmax classifier) of M neurons.

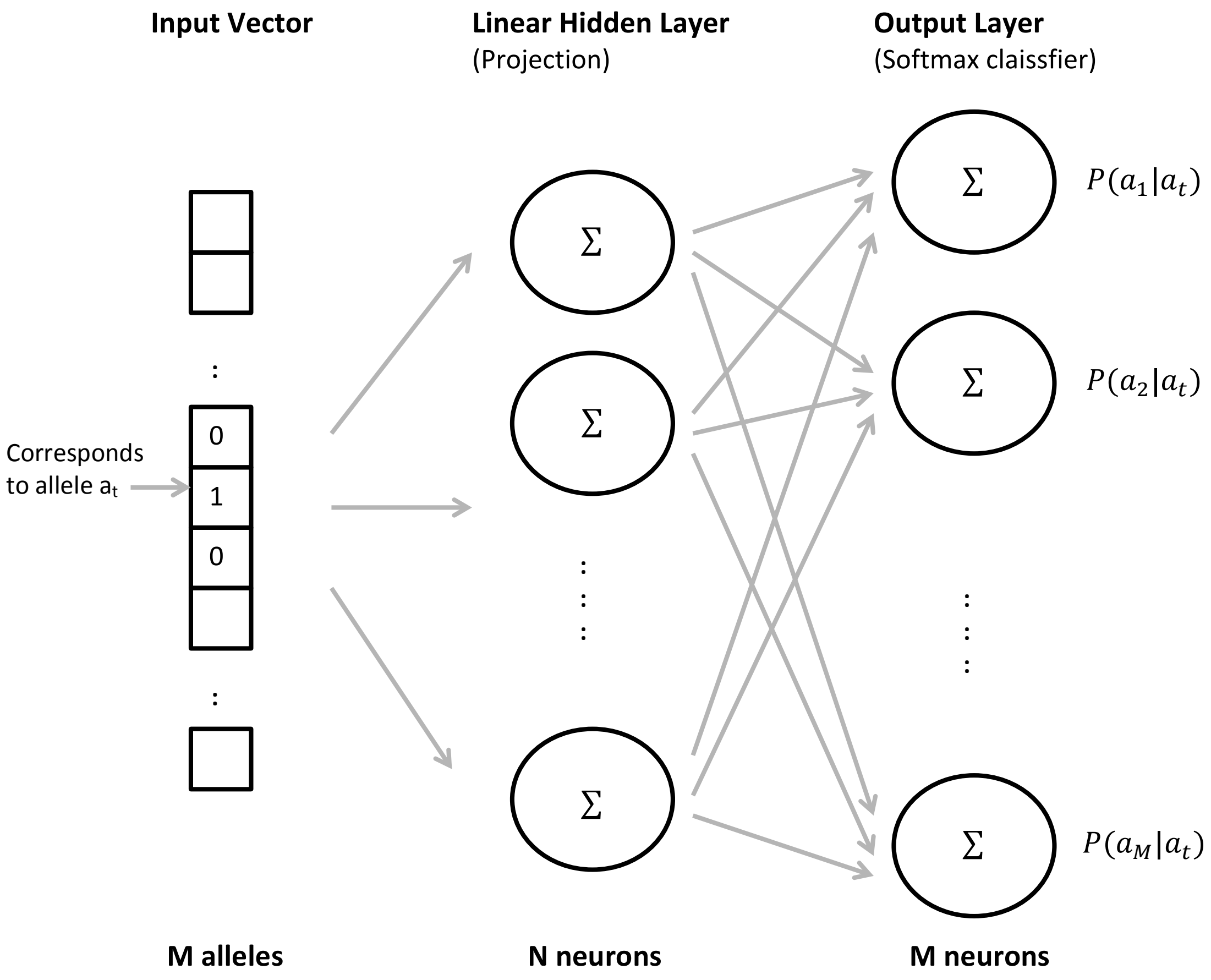

**Figure 2.** Workflow of Neural Network training from TensorBoard; nodes represent operations, solid lines represent data flow, and dotted lines represent control dependence edges.

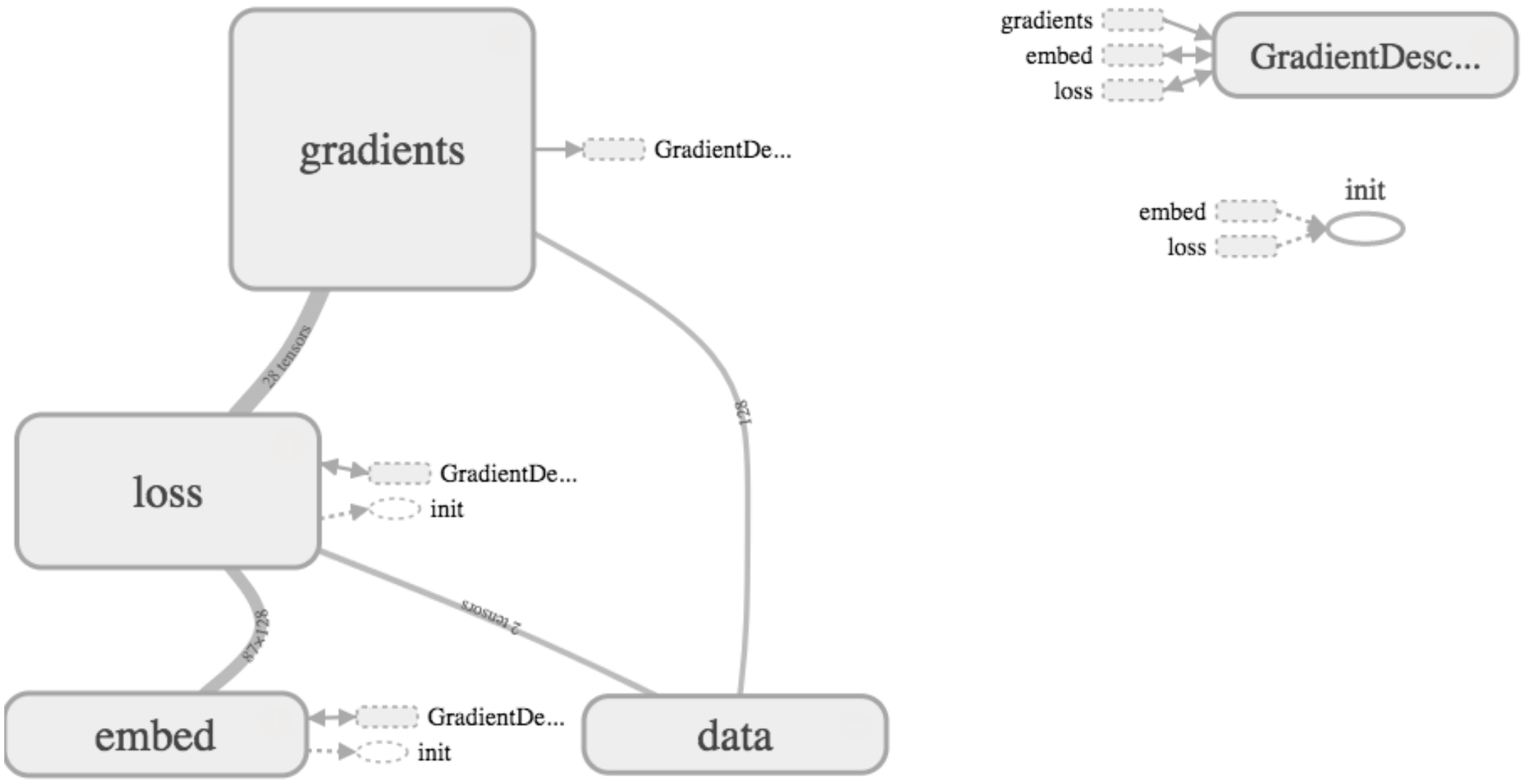

**Figure 3.** Pairwise correlation map of the first three PCA components of the allele embeddings from TensorBoard with $w=M/4$. The alleles are arranged in the genomic order: A. Exposed and B. Non-Exposed, and in the hierarchical clustering of seven clusters: C. Exposed and D. Non-Exposed. The green boxes indicate the cluster containing the mutation of interest, in the exposed population (P129Q denoted as Position 2844) and in the non-exposed population (H215N denoted as Position 3101).

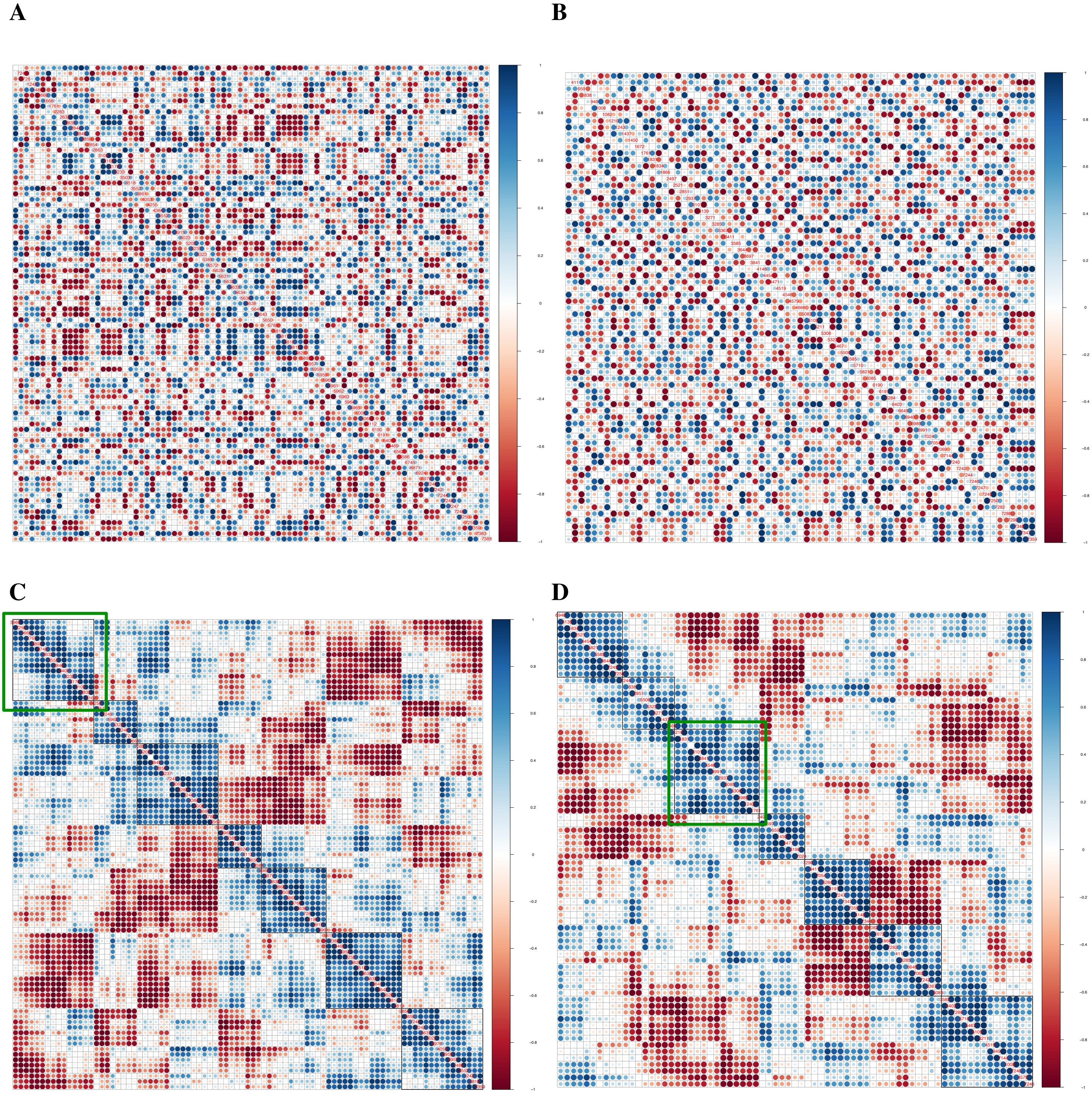

**Figure 4.** Non-synonymous mutations on the echovirus structural proteins (VP1: yellow, VP2: green, VP3: magenta, VP4: blue) identified as the cluster of potential adaptation to the disinfectant $ClO_2$ by the nucleotide skip-gran neural network. The images were generated by Chimera (Version 1.11.2) based on PDB entry 1H8T.

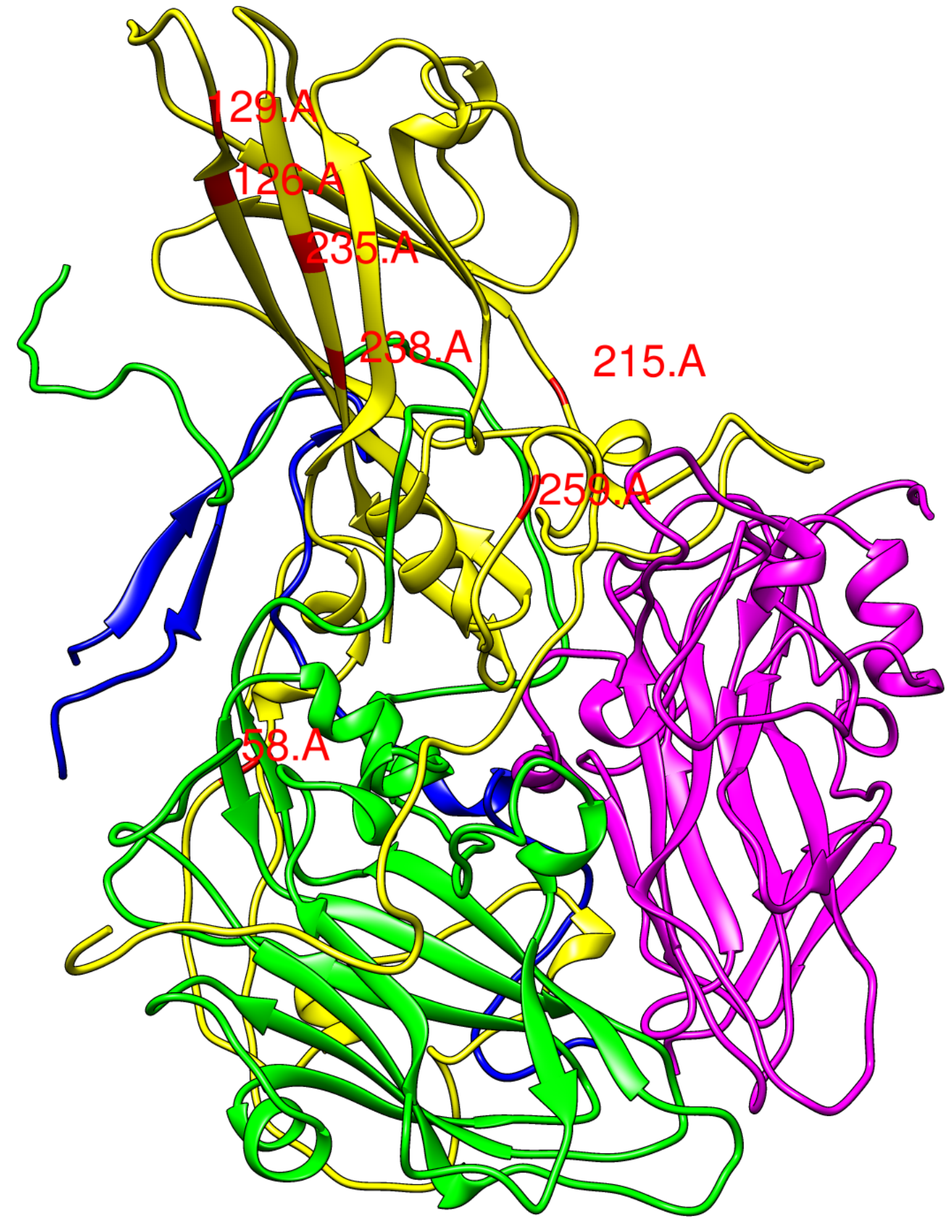